\documentclass[prl,twocolumn,preprintnumbers,amsmath,amssymb,floatfix]{revtex4-1}
\usepackage{graphicx}
\usepackage{epstopdf}
\usepackage{float}
\usepackage{placeins}
\usepackage{hyperref,graphicx}
\usepackage{dcolumn}
\usepackage{bm}
\usepackage{color}
\usepackage{xcolor}
\usepackage{natbib}
\usepackage{amsmath}
\usepackage{amssymb}
\usepackage{multirow} 
\usepackage{hyperref}
\usepackage{stackengine}
\usepackage{comment}
\hypersetup{colorlinks=true, linkcolor=blue, citecolor=red, urlcolor=magenta, pdftitle={2DMOF}, pdfauthor={Saurabh Ghosh}}
\begin{document}
\begin{titlepage}
\thispagestyle{empty}

\vspace*{\fill}

\end{titlepage}
\clearpage
\begin{center}
\begin{large}
\end{large}
\end{center}

\title{Symmetry-electronic fingerprints reveal competing magnetic phases in two-dimensional materials}
\author{Addis Fuhr$^1$}
\email{fuhras@ornl.gov}
\author{Zachary R Fox$^2$}
\author{David Parker$^3$}
\author{Ayana Ghosh$^4$}
\email{ghosha@iitm.ac.in}
\affiliation{$^1$Center for Nanophase Materials Sciences, Oak Ridge National Laboratory, Oak Ridge, TN, 37831, USA}
\affiliation{$^2$Computational Sciences \& Engineering Division, Oak Ridge National Laboratory, Oak Ridge, TN, 37831, USA}
\affiliation{$^3$Materials Science and Technology Division, Oak Ridge National Laboratory, Oak Ridge, TN, 37831, USA}
\affiliation{$^4$Department of Physics, Indian Institute of Technology, Madras, Chennai, 600036, India}

\begin{abstract}
Two-dimensional magnets offer compelling platforms for spintronics and quantum technologies, yet predicting their magnetic ground states, moments, and anisotropy remains challenging. This limitation primarily arises because existing machine-learning representations encode chemical environments without capturing the symmetry or exchange physics that govern magnetism. In this work, we introduce the symmetry-electronic fingerprint (SEF), a physically interpretable representation that encodes crystallographic symmetry operations, Wyckoff-site geometry, together with site-resolved electronic structure. Combined with ensemble learning with random forests, the SEF accurately classifies magnetic ordering while regressing moments alongside anisotropy energies while simultaneously resolving the distinct regimes of itinerant Stoner ferromagnetism from localized superexchange. What sets the SEF-trained models apart is that regions of elevated model uncertainty are not a failure but a diagnostic, identifying materials where these mechanisms compete. First-principles calculations on Co- and Ni-based halides and oxides confirm that these regions correspond to genuine near-degenerate FM and AFM phases with magnetic frustration, suppressed anisotropy, and emergent non-collinear ordering. By encoding symmetry together with exchange physics directly into the representation unlike conventional descriptors, the SEF transforms model uncertainty into a compass pointing toward two-dimensional materials where small perturbations drive transitions between collinear, frustrated, or non-collinear magnetic phases.
\end{abstract}

\maketitle
\section{Introduction}
Magnetism in two-dimensional (2D) materials has opened new opportunities for spintronics, information storage, and quantum technologies~\cite{gong2019two,gibertini2019magnetic}. Unlike their bulk counterparts, 2D magnets are inherently constrained by reduced dimensionality as well as symmetry, rendering long-range magnetic order fragile in the absence of strong magnetic anisotropy and spin-orbit coupling (SOC). Consequently, key magnetic properties such as ordering, moments, and magnetocrystalline anisotropy energy (MCAE) emerge from a subtle interplay of lattice symmetry, electronic structure, with spin degrees of freedom. Therefore, accurately capturing these coupled effects, together with the physical mechanisms behind them, remains a central challenge.
High-throughput density functional theory (DFT) has played a foundational role in identifying candidate 2D magnetic materials and deriving the mechanisms that stabilize magnetism in reduced dimensions. Large-scale computational screenings established the critical roles of magnetocrystalline anisotropy, local electronic correlations in addition to SOC in enabling finite-temperature magnetic order~\cite{Torelli2019_2DMater}, 
while subsequent studies expanded the chemical search space across broad families of van der Waals magnets~\cite{Rhone2020_SciRep}. Yet the computational cost of DFT severely limits exploration of complex chemical 
substitutions, multilayer configurations, and symmetry variations, motivating the development of data-driven alternatives.
%
\begin{figure*}
\centering
\includegraphics[width=0.9\textwidth]{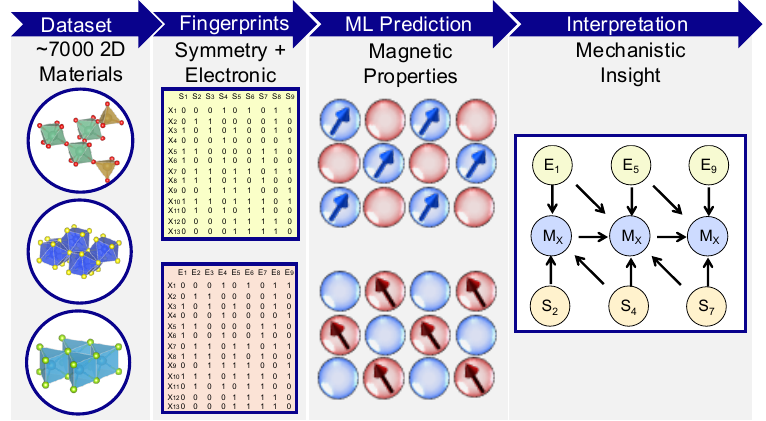}
\caption{Key steps in predicting and understanding magnetic properties of two-dimensional materials using the symmetry-electronic fingerprints. The fingerprint matrices encode the presence or absence of crystallographic symmetry operations (top) and elemental electronic descriptors (bottom) for each structure. The structures in the left panel (top to bottom) are Nb$_2$W$_2$O$_{11}$, TaS$_3$, TiSe$_2$.}
\label{fig:outline}
\end{figure*}
%

%
Machine learning (ML) approaches have emerged as powerful complements to DFT for magnetic materials discovery~\cite{kaba2023prediction,ghosh2020machine,horton2019high}. Supervised models trained on DFT-generated datasets have been applied to classify magnetic ordering, predict magnetic moments, and estimate anisotropy trends across diverse materials spaces~\cite{Xie2021_JPCL,Acosta2022_ACSAMI}. Interpretable ML approaches, 
including random forest (RF) models augmented with feature attribution techniques, have enabled the construction of magnetic material maps that can rationalize ferromagnetic versus antiferromagnetic behavior in terms of electronic and structural features~\cite{Acosta2022_ACSAMI}. More recently, efforts have focused on predicting magnetic moments in transition-metal-based 2D materials~\cite{singh2025simple}. Deep learning architectures such as graph neural networks have also been introduced to better capture structural connectivity and symmetry across diverse 2D materials~\cite{Elrashidy2024_JPCC, chotrattanapituk2026universal}.
Despite these advances, most existing input representations to ML models remain fundamentally atom-centered or graph-based, encoding local chemical environments without explicitly capturing the hierarchy of symmetry breaking that governs magnetic interactions. Magnetism is inherently tied to spatial symmetry, time-reversal symmetry, and their breaking through spin polarization, SOC, Dzyaloshinskii–Moriya interactions (DMI), 
and exchange pathways~\cite{litvin2008tables,bruno1989tight,dzyaloshinsky1958thermodynamic,
anderson1950antiferromagnetism,goodenough1955theory,kanamori1959superexchange}. Conventional descriptors whether based on local atomic environments, composition, or structural connectivity often succeed at prediction while 
obscuring the underlying physical mechanisms. In particular, they can neither distinguish between localized and itinerant magnetic regimes, nor expose competing interactions that suppress or enhance 
anisotropy~\cite{oviedo2022interpretable,zhong2022explainable}. As a result, materials poised at the boundary between competing magnetic phases remain effectively invisible to conventional screening.
Here we introduce the symmetry-electronic fingerprint (SEF), a unified descriptor that explicitly encodes crystallographic symmetry operations, Wyckoff-site geometry, coordination environments, together with site-resolved electronic structure. Rather than treating symmetry implicitly, the SEF is constructed to reflect the fundamental symmetry constraints governing SOC alongside exchange interactions. Combined with ensemble learning, the SEF enables physically meaningful structure-property relationships to be learned directly from these symmetry-electronic degrees of freedom, accurately classifying FM versus NM monolayers while regressing magnetic moments with anisotropy energies.
A key advantage of the SEF-trained ML models is that regions of elevated \textit{predictive uncertainty} are not a failure of the model but a \textit{diagnostic}. The mixed regions identified by self-organizing maps (SOMs)~\cite{kohonen2012self} of the SEF feature space correspond precisely to compounds where itinerant Stoner-like exchange competes with localized superexchange, identifying materials with near-degenerate FM, AFM, or frustrated states that are otherwise indistinguishable from ordinary magnets by conventional descriptors. Because symmetry and exchange physics are embedded directly into the representation, uncertainty in the SEF-trained models naturally points toward materials where collinear, frustrated, or noncollinear order compete on equal footing.
\section{Methodology}
\begin{figure*}
\centering
\includegraphics[width=0.8\textwidth]{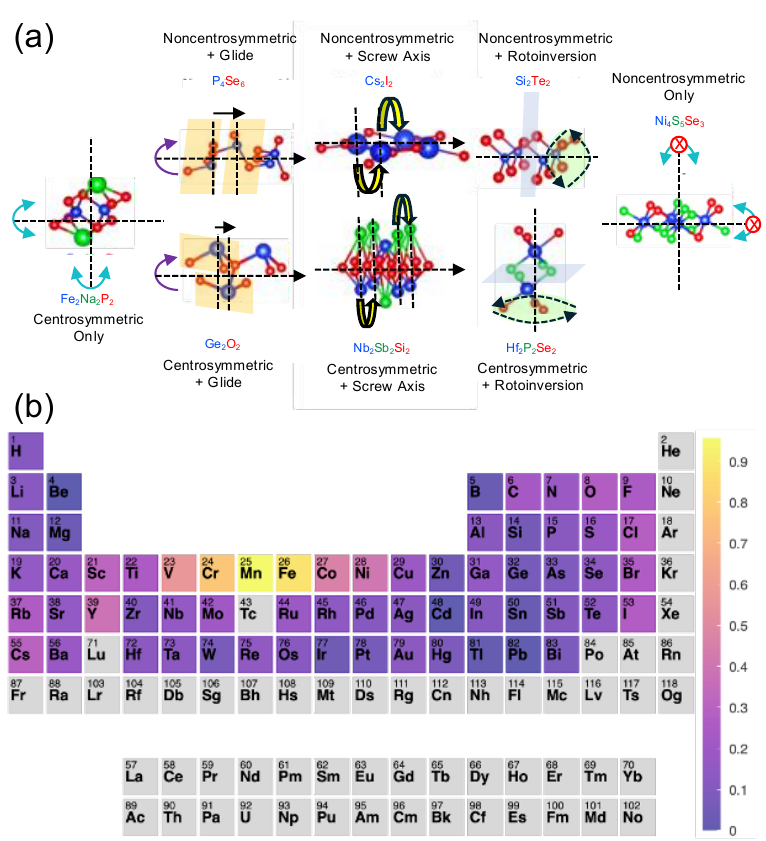}
\caption{Symmetry (a) and electronic (b) fingerprints. The major symmetry operations are shown for the symmetry fingerprint with example structures. The color contour for the electronic fingerprints represents the proportion of materials with those represented elements that are ferromagnetic, where grey indicates the element that is not represented in the dataset.}
\label{fig:SEF}
\end{figure*}

The key steps of our approach are illustrated in Figure~\ref{fig:outline}. 
Random forest (RF) ensemble models are trained to map SEF onto magnetic properties including magnetic moment, ordering type, anisotropy energy. Self-organizing maps (SOMs) are constructed to interpret the results to jointly visualize the learned feature space alongside predictive uncertainty. Regions of elevated uncertainty naturally highlight systems with competing magnetic interactions, guiding targeted first-principles calculations.
The SEF integrates crystallographic symmetry operations, Wyckoff-specific site symmetries, together with site-resolved electronic descriptors into a single invariant mathematical representation, illustrated for representative structures in Figure~\ref{fig:SEF}(a). By explicitly encoding these symmetry operations, the SEF captures how magnetic behavior is modulated by the local bonding environment. Electronic descriptors, referenced to atomic-level Wyckoff information, ensure the fingerprint reflects the key energetic and orbital factors governing moment formation and magnetic anisotropy. Aggregating site-resolved properties into composition-level statistics preserves chemical specificity, enabling models to learn how global lattice symmetries relate to electronic behavior, as illustrated in Figure~\ref{fig:SEF}(b).
%
%

\subsection{Mathematical Formulation of the Symmetry-Electronic Fingerprint (SEF)}
The SEF defines a mapping from the structural-symmetry and electronic state spaces of a crystal into a fixed-dimensional real-valued vector:
\begin{equation}
\Phi : (\mathcal{S}, \mathcal{E}) \rightarrow \mathbb{R}^n,
\end{equation}
where $\mathcal{S}$ is the structural-symmetry manifold defined by the crystallographic space group $G$, and $\mathcal{E}$ represents the electronic state space encoding elemental and orbital degrees of freedom.
Each crystal structure is decomposed into its fundamental symmetry operations:
\begin{equation}
\mathcal{O}_k \in G = \{ E, I, \bar{n}, S_n, \sigma_g, t|R \},
\end{equation}
where $E$ is the identity, $I$ denotes inversion, $\bar{n}$ rotoinversion, $S_n$ screw axes, $\sigma_g$ glide planes, and $t|R$ general Seitz operators combining translation $t$ with rotation $R$. Each operation $\mathcal{O}_k$ is encoded as a binary one-hot vector and concatenated into the symmetry component of the fingerprint:
\begin{equation}
\mathbf{s}_k =
\begin{cases}
1, & \mathcal{O}_k \in G,\\
0, & \text{otherwise},
\end{cases}
\quad
\Phi_{\text{sym}} = \bigoplus_k \mathbf{s}_k.
\end{equation}
Each atomic site $i$ is assigned a Wyckoff position and site symmetry subgroup:
\begin{equation}
w_i = (m_i, \mathbf{r}_i), \quad H_i \subseteq G,
\end{equation}
where $m_i$ is the site multiplicity, $\mathbf{r}_i$ the reduced coordinates, and $H_i$ the subgroup of operations leaving site $i$ 
invariant. The local coordination environment is described by a neighbor set within a cutoff radius $r_c$:
\begin{equation}
\mathcal{N}(i) = \{ j : \|\mathbf{R}_i - \mathbf{R}_j\| < r_c \}, \quad
d_{ij} = \|\mathbf{R}_i - \mathbf{R}_j\|,
\end{equation}
from which coordination statistics are computed:
\begin{equation}
C_\text{eff}(i) = \sum_{j \in \mathcal{N}(i)} f_\text{weight}(d_{ij}),
\qquad
\{\bar{d},\ \tilde{d},\ d_\text{mode},\ \Delta d\}_{ij},
\end{equation}
where $\bar{d}$, $\tilde{d}$, $d_\text{mode}$, and $\Delta d$ denote the mean, median, mode, and range of bond lengths, respectively.
Each atomic site $i$ is further associated with a vector of electronic attributes:
\begin{equation}
f_i = \{ Z_i,\ \chi_i,\ r_i^{\text{atom}},\ r_i^{\text{ion}},\ m_i,\  
n_i^{\text{val}},\ \text{IE}_i,\ \text{EA}_i,\ E_{g,i} \},
\end{equation}
referenced to its Wyckoff site $w_i$ and site symmetry $H_i$.
Composition-level electronic features are obtained by aggregation:
\begin{equation}
\Phi_{\text{elec}} = \text{Agg}_{i \in \text{sites}}(f_i),
\end{equation}
where $\text{Agg}$ computes the mean, median, range, and mode of each 
attribute across all sites. The full SEF is formed by concatenating the 
symmetry and electronic components:
\begin{equation}
\Phi = \Phi_{\text{sym}} \oplus \Phi_{\text{elec}} \in \mathbb{R}^n.
\end{equation}
%
By construction, the fingerprint satisfies three invariance conditions 
essential for a physically meaningful representation:
\begin{align}
\Phi(P \mathbf{R}) &= \Phi(\mathbf{R}), 
& &\text{(permutation of identical atoms),}\\
\Phi(R \mathbf{R} + t) &= \Phi(\mathbf{R}), 
& &\text{(rotation and translation),}\\
\Phi(\mathcal{O}_k \mathbf{R}) &= \Phi(\mathbf{R}), 
& &\text{(crystallographic symmetry operations).}
\end{align}
The first condition ensures the fingerprint is independent of the arbitrary labeling of equivalent atoms. The second guarantees invariance to rigid-body transformations, making the representation transferable across unit-cell choices. The third enforces consistency with the full space-group symmetry of the crystal, ensuring that symmetry-equivalent structures map to identical fingerprints. Together, these conditions yield a symmetry-complete, physically interpretable embedding that captures both the allowed magnetic interactions and the influence of electronic structure on their strength. Implementation details are provided in the Supplementary Information.
\subsection{Data Curation}
Monolayer unit cell structures are sourced from the Computational 2D 
Materials Database (C2DB)~\cite{haastrup2018computational}. For each 
entry, we extract the relaxed atomic geometry alongside the reported 
magnetic state (NM, FM, or AFM), total magnetic moment, magnetic 
anisotropy energies ($E_{\text{zx}}$ and $E_{\text{zy}}$), and band 
gap $E_g$; entries missing any of these quantities are removed.

Crystallographic symmetry descriptors are generated using \texttt{spglib} \cite{togo2024spglib} 
and \texttt{pymatgen}\cite{ong2013python}. Each structure is reconstructed from the ASE 
\texttt{Atoms} object, from which the space group (Hermann-Mauguin 
symbol and number) and Wyckoff positions are determined. For every 
atomic site, nearest neighbors are identified using a 3.3~\AA\ cutoff, 
from which coordination numbers, bond-length distributions, and 
site-multiplicity statistics are computed. This cutoff is chosen to 
encompass covalent, ionic, metallic, chalcogenide, and oxide bonding 
environments without capturing second-shell distances for most structures. 
Mean, median, mode, and range are evaluated for each structural descriptor. 
Only the mean is retained for effective coordination number and site 
multiplicity. Structures with no neighbors within the cutoff are excluded.
The in-plane area per atom is included as a coarse geometric-density 
descriptor.

Space groups are categorized according to the 
presence~\cite{sands1993introduction} of inversion centers, screw axes, 
glide planes, and roto-inversion operations, each one-hot encoded. Wyckoff 
positions and atomic numbers are likewise one-hot encoded to preserve 
discrete site-specific information; their joint occurrence yields the 
full SEF representation. A complete one-hot encoding of space-group 
number is retained for interpretability analyses but excluded from model 
training. For auxiliary analysis, stoichiometry-weighted elemental 
descriptors such as electronegativity, electron affinity, valence electrons, 
ionization energy, atomic mass, and atomic radius are summarized 
using mean, median, mode, and range. Magnetic states are encoded 
numerically (NM~$= 0$, FM~$= 1$, AFM~$= 2$); for NM structures, the 
total magnetic moment is set to zero when not explicitly reported. Band 
gaps are taken directly from the database in eV.
\subsection{ML Models}
RF models are employed throughout owing to their ability to handle 
heterogeneous feature types, resistance to overfitting on modest sample 
sizes, and direct interpretability through feature-importance scores. These are 
well-suited to the mixed discrete-continuous structure of the 
SEF. The descriptor set augments the SEF with the electronic band gap as 
a continuous feature and coordination-derived statistics, together 
capturing the contrast between Stoner-like metallic exchange and 
superexchange in finite-gap 
insulators~\cite{litvin2008tables,bruno1989tight,dzyaloshinsky1958thermodynamic,
anderson1950antiferromagnetism,goodenough1955theory,kanamori1959superexchange}.

\begin{figure*}
\centering
\includegraphics[width=\textwidth]{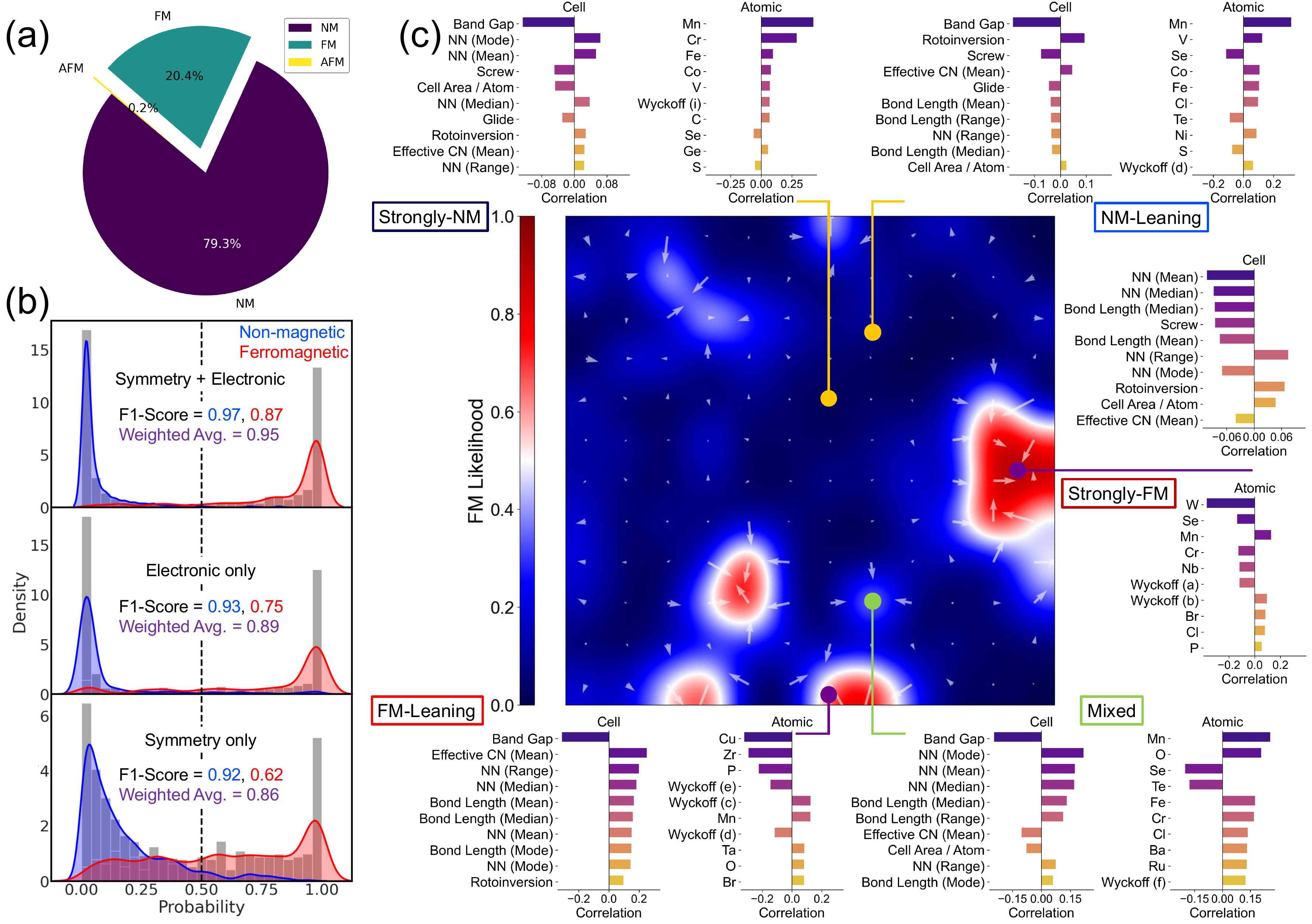}
\caption{Distribution of magnetic states in the C2DB database (a). The classification task is limited to the FM and NM states, and the results for symmetry only, electronic only, and combined SEF fingerprints are shown in (b). A SOM map is generated (c) to divide the feature space into regions based on FM likelihood while drawing insights on the representation-based uncertainty.}
\label{fig:ML_model1}
\end{figure*}

For magnetic-state classification, a binary RF classifier distinguishes 
FM from NM monolayers (Figure~\ref{fig:ML_model1}(a)). Since NM entries 
dominate the dataset, class imbalance is addressed by oversampling the 
minority FM class via sampling with replacement until both classes reach 
equal cardinality. A randomized tree ensemble with class weights inversely 
proportional to class frequencies further penalizes FM misclassifications 
without modifying the underlying data distribution.

For regression, only FM materials are considered. The targets are the 
absolute magnetic moment per atom ($\mu_B$/atom) and the anisotropy 
energies $E_{\text{zx}}$ and $E_{\text{zy}}$ (meV/atom), both of which 
follow strongly skewed distributions with many small values and few large 
outliers. A bin-and-resample strategy is applied to the training set to 
mitigate this imbalance. Magnetic moments are partitioned into four bins: 
$0 < \mu_B/\text{atom} \leq 0.5$, $0.5 < \mu_B/\text{atom} \leq 1.0$, 
$1.0 < \mu_B/\text{atom} \leq 1.5$, and $\mu_B/\text{atom} > 1.5$. 
MCAE values are binned by absolute magnitude: 
$10^{-6} < |\text{MCAE}| \leq 2.5\times10^{-2}$, 
$2.5\times10^{-2} < |\text{MCAE}| \leq 2.5\times10^{-1}$, 
$2.5\times10^{-1} < |\text{MCAE}| \leq 2.5$, and $|\text{MCAE}| > 2.5$ 
(all in meV/atom). Values below $10^{-6}$~meV/atom are discarded as 
numerical artifacts. The lowest retained bin captures near-zero MCAE 
within the practical accuracy limit of DFT SOC calculations. Each bin 
is upsampled via bootstrap resampling to match the largest bin, producing 
a balanced training set while preserving the original regression targets.

All models use an 80/20 train-test split with random shuffling. 
Hyperparameters such as tree count, maximum depth, and minimum samples per 
leaf are optimized by grid search with 10-fold cross-validation. 
Model performance is quantified using accuracy, balanced accuracy, and 
receiver operating characteristic (ROC) 
metrics~\cite{breiman2001random}. Feature-importance scores from the 
trained RF models identify which SEF components most strongly govern 
magnetic behavior, forming the basis for the mechanistic analysis in 
the following section.
\subsection{First-Principles Simulations}
First-principles calculations were performed within density functional 
theory (DFT) using the projector augmented-wave method as implemented 
in VASP \cite{kresse1996efficiency, kresse1996efficient} with the PBE exchange-correlation functional. Selected 
two-dimensional materials from the C2DB database were modeled using 
$4 \times 4 \times 1$ supercells. Structural relaxations were carried 
out across multiple magnetic configurations, for e.g., NM, FM, and 
symmetry-distinct AFM states to identify magnetic ground states.

MCAEs were computed using noncollinear DFT with SOC, 
while fixed-spin-moment calculations were employed to assess itinerant 
versus localized magnetic behavior. Magnetic exchange interactions up 
to third-nearest neighbors were extracted by mapping DFT total energies 
onto a Heisenberg model, providing a first-principles parameterization 
of exchange couplings. These parameters were subsequently used within 
the Luttinger-Tisza formalism to determine magnetic ordering vectors and 
identify candidate collinear and noncollinear spin states. Further 
computational details including convergence parameters, magnetic 
configurations, exchange mapping procedures, and Luttinger-Tisza 
analysis are provided in the Supplementary Information.
%
\section{Results \& Discussion}
\subsection{Classifying magnetic order with SEF}
The performance of the RF classifier model trained to distinguish between FM and NM classes is shown in Figure~\ref{fig:ML_model1}(b) and Table~\ref{tab:perf}. 
%
\begin{table}[]
    \centering
    \begin{tabular}{c|c|c|c}
        Magnetic State & Precision & Recall & F1 Score \\
        NM & 0.97 & 0.97 & 0.97 \\
        FM & 0.89 & 0.87 & 0.88
    \end{tabular}
    \caption{Performance metrics of the classification model to categorize between FM and NM classes using SEF.}
    \label{tab:perf}
\end{table}
%
Figure~\ref{fig:ML_model1} includes the individual class and class balanced accuracy, precision, recall, and F1 scores of models trained with with symmetry fingerprints only, electronic fingerprints only, and the SEF, respectively. 
These metrics allow us to quantify both overall classification success and the specific distribution of false positives and false negatives between NM/FM categories. 
F1 scores provide a rigorous measure of overall model performance, as they capture the harmonic mean of precision and recall, thereby balancing the trade-off between false positives and false negatives. 
As expected, the inherent class imbalance results in systematically higher performance for the NM class compared to FM, respectively. 
Such disparity, however, is progressively mitigated as richer feature sets are employed. 
When trained with symmetry-based descriptors alone, the F1 score for FM is $\sim$30\% lower than that for NM.
The gap narrows to $\sim$20\% when using electronic fingerprints alone.
The difference is further reduced to only $\sim$9\% when SEF is employed. 
Integrating these complementary fingerprints yields a weighted average F1 score of 0.95, underscoring the ability of the unified representation to capture the fundamental physics governing magnetism in 2D materials.
To further elucidate the structure-property relationships underlying magnetic interactions, we map the SEF components onto distinct magnetic classes. 
We first address the features that separate FM from NM 2D materials by applying SOMs to the output of our RF classifier, thereby enabling unsupervised learning to perform segmentation of the RF learned feature space.
This is illustrated in Figure~\ref{fig:ML_model1}(c).
SOMs (Kohonen artificial neural networks) nonlinearly embed high-dimensional input into a low-dimensional space via competitive learning with a neighborhood function, yielding a discretized representation (“map”) that approximately preserves topological relationships. 
Unlike clustering, SOMs do not assign classes. 
Rather, it finds proximity in fingerprint similarity in the original feature space. 
After SOM training is completed, each sample is mapped to its best-matching unit (BMU) on the map, enabling visualization of label distributions. Here, we focus on the classification of magnetic ordering.
To quantify the propensity of certain SEF components to yield FM materials, we calculate the FM likelihood for each SOM region, which is defined as the fraction of mapped samples predicted to be FM. 
A value of 1 denotes that all samples in the region are FM, whereas 0 indicates none are FM. 
Overlaid on this surface is a white vector field representing the local gradient highlighting pathways along the SOM regions where FM likelihood is increasing, and delineates transition corridors linking NM-dominated and FM-dominated basins. 
For downstream interpretability, we partition the landscape into five regimes: Strong FM (0.9-1), FM-Leaning (0.75-0.9), Mixed (0.25-0.75), NM-Leaning (0.1-0.25), and Strong NM (0-0.1). 
The Pearson correlation coefficient (r) is computed for each region by comparing every symmetry and electronic fingerprint to the FM label. 
Feature ranking is based on the magnitude of r, or $|r|$ indicating the SOM region specific importance, and the sign of r representing directionality. 
Specifically, positive value for r indicates the feature is associated with the FM phase, a negative value the NM phase, and the magnitude of the sign its importance in either direction. This yields compact, initial regime-resolved structure-property associations.
\subsection{Symmetry-resolved magnetic interactions with SOMs}
The trends captured by the SOMs enable us to identify global and region-specific mechanistic drivers of magnetism while pointing at two-dimensional systems with particularly interesting magnetic characteristics. 
Band gap (electronic fingerprint) exhibits a highly robust negative correlation with FM propensity. 
This observation points to a pronounced Stoner-type mechanism, wherein itinerant FM arises from an enhanced density of states (DOS) near the Fermi level (E$_F$). 
The resulting increase in the product of the Stoner exchange integral \(I\) and the density of states at the Fermi level \(N(E_F)\) drives the system toward the instability criterion \(I \cdot N(E_F) > 1\), marking the transition at which a nominally nonmagnetic metal spontaneously develops ferromagnetic order through exchange interactions among itinerant electrons.
For 2D materials, this implies that metallic, semi-metallic, or small band gap materials with partially filled bands are predisposed toward FM ordering. 
In contrast, large band gap structures tend to stabilize either closed-shell NM states or AFM configurations governed by super-exchange interactions while the latter is not explicitly represented in the dataset as used for RF classification.
Early 4$d$/5$d$ cations (e.g., Zr, Hf, and W, electronic fingerprints) are also correlated with the strongly FM region of the SOM map, while the NM-leaning region is dominated by Se, Te, and O anions. 
This suggests an underlying mechanism consistent with that described above. 
As expected from chemical intuition, the electronic bandwidth is broadened by the delocalized nature of 4$d$/5$d$ orbitals, which enhances N(E$_F$), strengthens itinerant ferromagnetism, and promotes double-exchange channels. We note, however, that this trend contrasts with experimental behavior, where early 4$d$/5$d$ transition metals more commonly exhibit AFM or NM ground states, and broader bandwidths generally reduce N(E$_F$). The FM tendency captured here likely reflects a systematic overestimation of spin polarization inherent to PBE, which is known to over-delocalize $d$ electrons in 4$d$/5$d$ systems. Since the SEF is trained on C2DB data computed consistently at the PBE level, the SOM faithfully reproduces these functional-level trends. For truly accurate prediction of ground-state physics, DFT calculations with a higher-level functional would be required. 
As established in prior studies \cite{burch2018magnetism,gong2019two,hou2025magnetic}, strong SOC is often required in 2D systems to provide sufficient magnetocrystalline anisotropy to overcome Mermin-Wagner fluctuations and sustain long-range magnetic order. 
Chalcogens and oxides, on the other hand, increase $p-d$ covalency. 
This, in turn, stabilizes superexchange as described by the Goodenough-Kanamori rules and is particularly important for edge-sharing octahedral geometries, thus driving magnetic order away from FM and toward NM or AFM configurations. 
This trend is further supported by specific Wyckoff motifs (symmetry fingerprints, e.g., W:b), which reflect coordination geometries correlated with the strongly FM region. 
Specifically, the contrast between 1H trigonal-prismatic and 1T octahedral environments illustrates how the sign and bandwidth of superexchange can be tuned via bond-angle changes from 90° to 180°. Hence, the correlation of the W:b geometry with the strongly FM region is consistent with the expected geometry-induced promotion of itinerant overlap via edge sharing and shortened interatomic distances between metal-metal bonds.
Larger coordination numbers and broader distributions of nearest-neighbor (NN) distances and bond lengths (BL), coming from the symmetry part of the SEF are characteristic of the FM-leaning and strongly FM regions.
Physically, an increase in coordination number expands the number of available exchange pathways per magnetic center, thereby amplifying the net exchange interaction. We note that the apparent FM specificity of this trend may partly reflect the limited representation of AFM configurations in the C2DB dataset, as discussed above, rather than a genuine preference for FM over AFM ordering. 
Meanwhile, broader NN and BL distributions reflect local anisotropy and structural flexibility, often associated with Peierls distortions, dimerization, or Jahn-Teller effects. Such distortions can lift orbital degeneracies to promote double-exchange or itinerant ferromagnetism, or alternatively, modulate the sign and magnitude of super-exchange by tuning $p-d$ hybridization. 
A modest positive correlation is also observed between larger in-plane area per atom (symmetry fingerprint) and FM-leaning behavior, suggesting that expanded lattice spacing weakens direct antiferromagnetic superexchange.
In turn, it reduces the AFM coupling constant more strongly than the FM contribution.
In the presence of itinerant carriers, it shifts the balance toward double-exchange or itinerant ferromagnetism. This trend is consistent with the well-known distance-dependent crossover between AFM and FM exchange observed in transition metals such as Mn, where increasing interatomic separation drives a transition from AFM to FM alignment, as described by the Bethe-Slater curve.
Centrosymmetry and nonsymmorphic operations (glide planes, screw axes, and rotoinversion) as captured in SEF exhibit weak but discernible correlations with FM-leaning and mixed regions. 
Specifically, centrosymmetry shows a positive correlation with FM-leaning behavior, whereas nonsymmorphic operations display a negative one. 
These correlations become more pronounced for glide planes, screw axes, and rotoinversion symmetries as the feature space shifts toward NM-leaning and strongly NM regions. 
Together, this suggests that nonsymmorphic symmetries protect nodal crossings, band degeneracies that persist unless lifted by SOC. 
Specifically, SOC broadens the bands near the Fermi level, reducing N($E_F)$ and thereby weakening the Stoner instability criterion, pushing materials toward nonmagnetic behavior. In systems with nonsymmorphic symmetry, this effect is further compounded as SOC can flatten or open protected nodal crossings, additionally reducing spin polarization.
The presence or absence of centrosymmetry further determines whether Dzyaloshinskii–Moriya (DM) interactions are symmetry-allowed. 
For instance, within SEF, noncentrosymmetric monolayers promote canted or chiral spin textures, whereas centrosymmetric counterparts can stabilize ferromagnetism by suppressing DM interactions.
Collectively, the SOM analysis reveals a coherent hierarchy of magnetic mechanisms across the feature space. 
Metallic, semimetallic, or narrow band gap 4$d$/5$d$ compounds exhibiting broad bond-length distributions and high effective coordination numbers tend to host itinerant, Stoner-type FM orderings. 
These systems can be further stabilized by mixed-valent double-exchange and by strong SOC-driven magnetocrystalline anisotropy, which preserves long-range order in 2D. The FM-leaning regime shares this itinerant character, but reflects geometry-dependent exchange: the \(I \cdot N(E_F) > 1\) product lies near the Stoner threshold, and reduced bandwidths allow fluctuations into non-FM states. 
In contrast, NM-leaning and strongly NM materials exhibit larger band gaps, pronounced nonsymmorphic symmetry operations (screw axes, glide planes, and/or rotoinversion), and chalcogenide or oxide chemistries consistent with superexchange-dominated or NM ground states.
%
\begin{figure*}
\centering
\includegraphics[width=\textwidth]{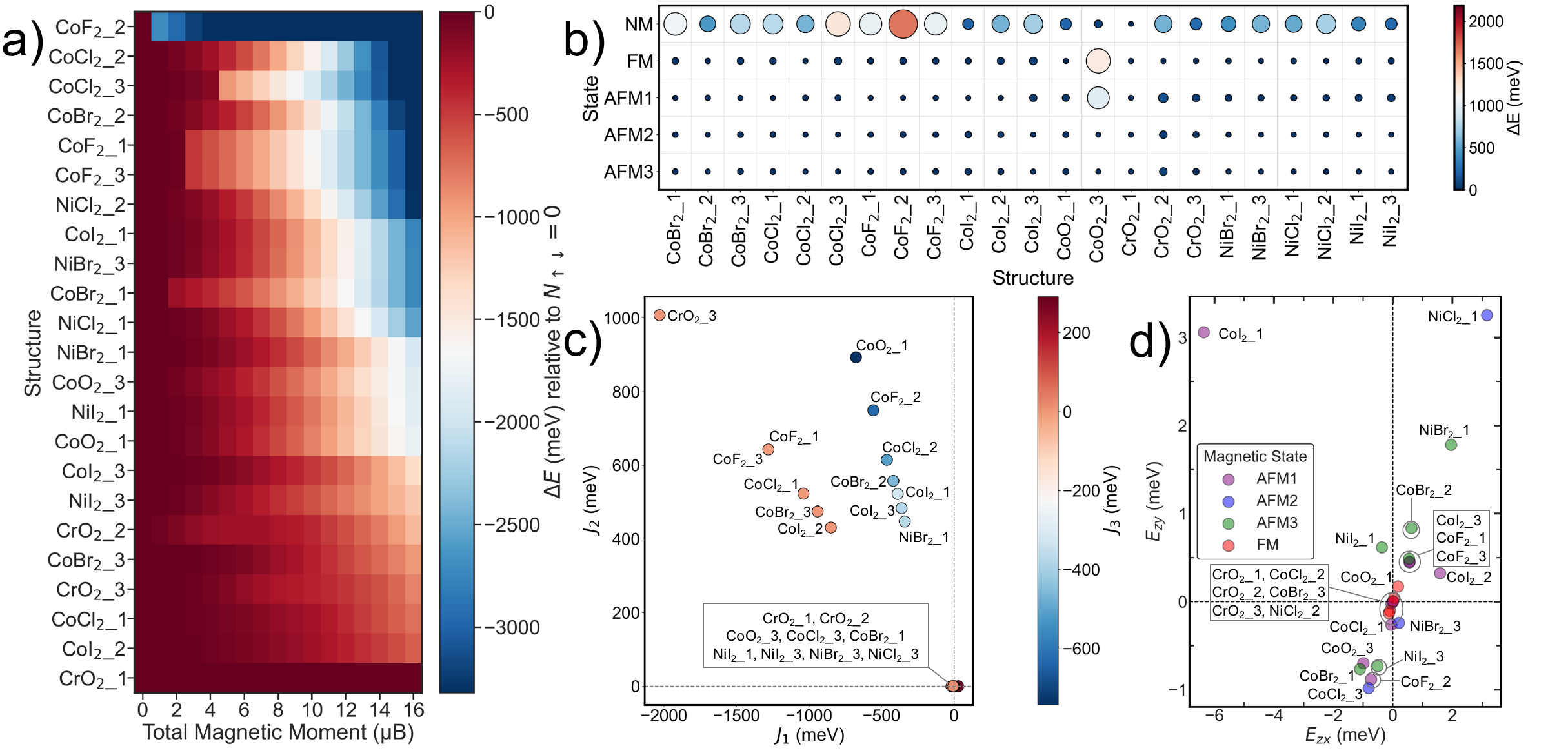}
\caption{(a) Relative total energies of ferromagnetic (FM) and antiferromagnetic (AFM1–AFM3) configurations referenced to the nonmagnetic (NM) state, illustrating the energetic proximity and competition among magnetic states for materials in the mixed SOM region. (b) DFT computed total-energy differences between FM, NM, and AFM (AFM1–AFM3) configurations, revealing near-degenerate magnetic states. (c) Exchange mapping onto a \(J_1\)–\(J_2\)–\(J_3\) Heisenberg model, highlighting competition between direct exchange and superexchange interactions. (d) Magnetocrystalline anisotropy energies from SOC calculations, distinguishing weakly anisotropic itinerant FM systems from strongly anisotropic localized AFM materials.}
\label{fig:simulations}
\end{figure*}
\subsection{Model uncertainty as a diagnostic for competing exchange} 
The \textit{mixed region}, characterized by elevated predictive uncertainty, exhibits coexisting itinerant and super-exchange features, including intermediate band gaps, coordination environments, and symmetry motifs that activate competing magnetic channels. 
Because our models are trained on SEFs that retain direct physical meaning, the uncertainty localized in these SOM regions is not incidental but instead reflects proximity to multiple, nearly degenerate magnetic mechanisms. 
Consequently, these regions serve as a targeted design space for ab initio exploration, where small perturbations in structure, composition, or strain may stabilize distinct magnetic ground states and give rise to phase competition or magnetic frustration in two-dimensional materials.
More broadly, this demonstrates that physically grounded representation learning plays a central role in materials discovery even when paired with non-parametric models such as RFs: by embedding mechanistic information into the feature space, model uncertainty becomes a diagnostic tool that highlights materials poised at the boundary between competing magnetic regimes, rather than a measure of prediction failure.
\subsection{From SEF predictions to first-principles magnetic ground states}
We therefore perform DFT calculations to test whether the SOM-identified mixed regime indeed represents a frontier of near-degenerate FM and AFM phases driven by competing exchange couplings (J$_\text{ij}$, Figure S2). 
Specifically, we probe itinerant Stoner instabilities and crossover behavior between localized and delocalized magnetic mechanisms
as shown in Figure~\ref{fig:simulations}.
Since the original dataset and SOM classifier excluded AFM configurations, this extension explicitly captures competing exchange channels predicted by the preceding mechanistic trends. 
Guided by both chemical intuition and SEF, we select representative compounds spanning Co- and Ni-halides, Co- and Ni-oxides, and Cr-oxides – chemistries that recurrently occupy the mixed region. 
Co and Ni halides exhibit reduced crystal-field splitting and moderately ionic bonding, leaving 3$d$ states partially delocalized and prone to itinerant or double-exchange ferromagnetism. 
Their oxide analogs display stronger covalency and electron correlation, favoring localized $d$-states and superexchange-dominated AFM or crossover behavior. 
Cr oxides, with half-filled 3$d^5$ shells and high covalency, are also explored for their potential approach of the super-exchange limit.
These structures also include multiple coordination motifs ($P-6m2$ or 1H, $P-3m1$ or 1T, and $P-4m2$) varied electronic structures such as metallic, half-metallic, semimetallic, and semiconducting (Figure S3).
These inclusions enable direct comparison between trigonal-prismatic and octahedral environments, probing the influence of crystal-field splitting, orbital overlap, and symmetry-dependent exchange. 
The $P-4m2$ structure adds a tetragonal distortion that lowers site symmetry and enables us to explore the possibility of anisotropic exchange, which is particularly relevant for SOC-driven magnetocrystalline effects. 
Collectively, these systems provide a controlled framework for examining how electronic structure, bandwidth modulation, exchange anisotropy, and geometric frustration mediate the interplay between itinerant and localized magnetism in 2D materials informed by SEFs.
To test such hypotheses outlined above, we aim to examine the degree of itinerancy in these representative systems.
We construct larger supercells of each structure – Co- and Ni-halides and oxides, as well as Cr-oxides – capable of accommodating multiple spin configurations (NM, FM, AFM1, AFM2, AFM3) for DFT calculations. 
From the NM starting point, we performed fixed-spin-moment (FSM) calculations by varying the total magnetic moment and mapping the total energy as a function of imposed spin polarization. 
Figure~\ref{fig:simulations}(a) presents the relative energy change with respect to the NM state (FSM = 0), while the absolute energies and corresponding DOS for the most stable configuration are shown in Figure S3.
The resulting FSM “Stoner maps” confirm that the SOM-predicted symmetry and electronic fingerprints correlate strongly with itinerant behavior: Co- and Ni-halides display metallic, semimetallic, or narrow-gap electronic structures (Figure S3) with pronounced DOS peaks near E$_F$, enhancing $I$ $\cdot$ N($E_F)$ and yielding a nearly monotonic energy decrease with increasing FSM. 
This continuous evolution of magnetization is characteristic of Stoner-type exchange rather than quantized, localized-moment transitions. 
Within the tested range (up to FSM = 16$\mu_B$ per supercell, which represents the number of transition metals in the supercell), no saturation of the total magnetic moment is observed, consistent with band-driven itinerant spin polarization. 
However, several counterintuitive trends point to competing magnetic mechanisms. 
In a simple Stoner framework, heavier and more polarizable halides (Br, I) are expected to enhance itinerancy through bandwidth narrowing and increased magnetic instability. 
Instead, the opposite trend emerges: the smaller and more ionic anion (F) produces the strongest stabilization and the largest total-energy reduction upon spin polarization. 
This inversion implies that shorter, more ionic metal-anion bonds amplify intra-atomic exchange and promote itinerant spin splitting.
Geometric effects further modulate this competition. For CoX$_2$ (where X = I, Br, or Cl), the octahedral 1T structure exhibits the strongest magnetic stabilization, followed by the trigonal-prismatic (1H) and the tetragonally distorted $P-4m2$ polytype, which remains comparatively localized.
This is consistent with larger crystal-field gaps and reduced N(E$_F$). 
CoO$_2$ follows a similar trend to the non-fluorinated halides, though the magnitude difference between 1H (semimetallic) and 1T (metallic) is smaller, reflecting reduced localization. 
Upon fluorination, this hierarchy inverts: trigonal-prismatic coordination compresses the $d-p$ manifold at E$_F$, requiring only minimal exchange splitting to induce a sharp, spin-selective drop in total energy. 
CrO$_2$ deviates markedly from these trends.
The 1H phase remains semiconducting and nonmagnetic, while the 1T and $P-3m1$ phases are half-metallic and exhibit “kinks” in the FSM curves (Figure S3), signaling a spin-channel crossover where one spin gap closes. 
Together, these observations highlight how crystal-field environment, bond ionicity, and electronic bandwidth collectively determine whether 2D materials exhibit itinerant, localized, or complex mixed exchange-driven magnetism.
To further probe the competing magnetic mechanisms in the mixed region of the SOM, we compared the DFT total energies of the ferromagnetic (FM) phase with those of the nonmagnetic (NM) state and three distinct antiferromagnetic configurations (AFM1–AFM3; Figure~\ref{fig:simulations}(b). 
Among the 23 structures examined, 18 favor an AFM ground state, while only five exhibit robust FM stability: CoCl$_2$ (1T), CoO$_2$ (1H), CrO$_2$ (1H and 1T), and NiCl$_2$ (1T). 
To assess the role of superexchange and its connection to the Goodenough–Kanamori rules, we performed magnetic exchange mapping (Figure~\ref{fig:simulations}(c) to identify competing interactions and potential spin frustration. 
Magnetic exchange parameters \(J_{ij}\) are extracted by mapping the total energies of the FM and AFM configurations onto a classical Heisenberg Hamiltonian (eqn. 13). Spin–spin correlations are evaluated for each coordination shell, and the magnitudes and signs of the exchange interactions (FM versus AFM) were determined through least-squares fitting to the DFT total energies. 
This procedure provides a first-principles parameterization of short-range Heisenberg couplings, enabling direct assessment of magnetic ordering tendencies and the emergence of spin frustration, an analysis that can be further refined through the inclusion of spin–orbit coupling (SOC) effects.

\begin{figure*}
\centering
\includegraphics[width=\textwidth]{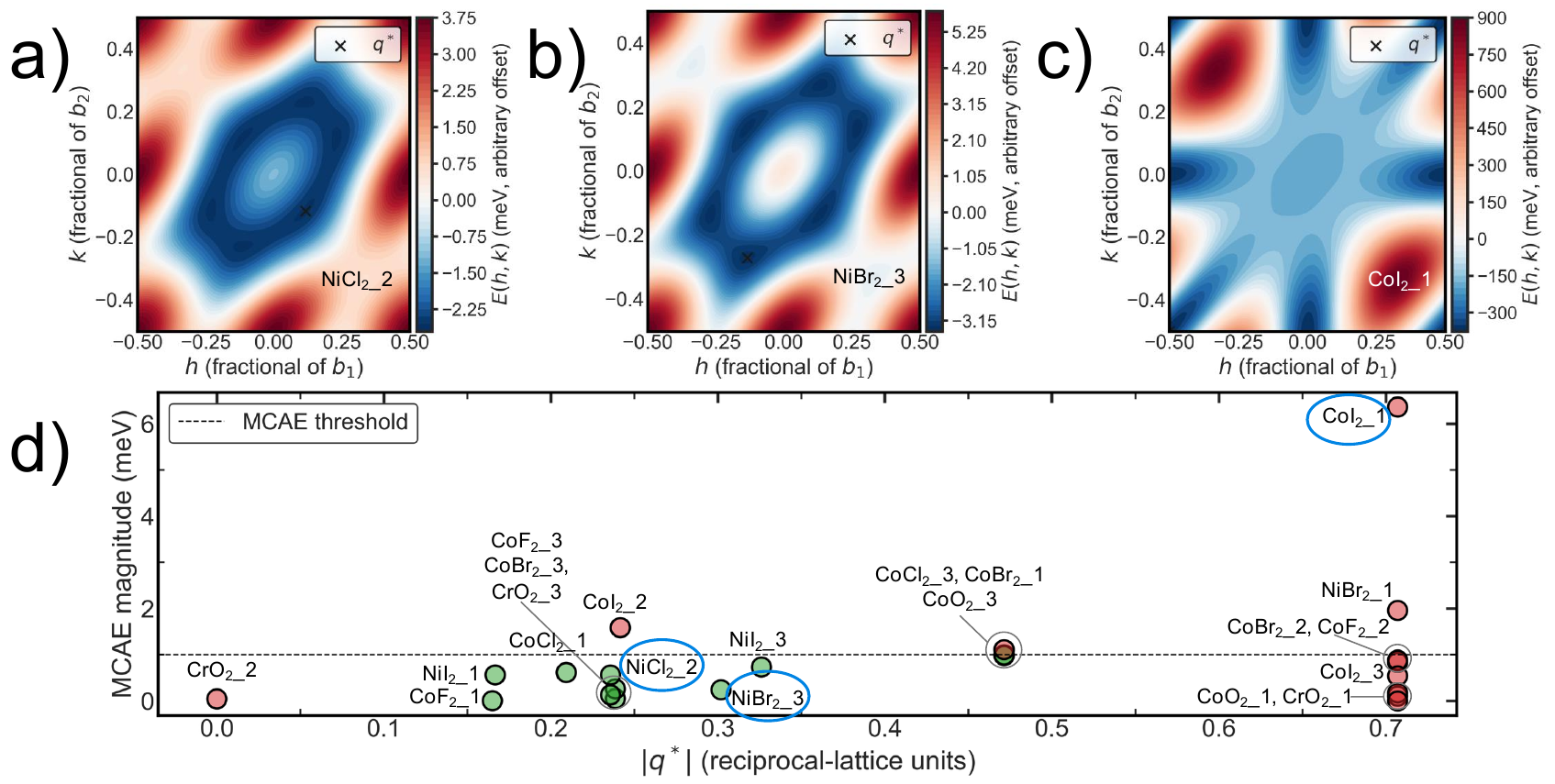}
\caption{(a-c) Luttinger–Tisza exchange-energy landscapes for NiCl$_2$, NiBr$_2$, and CoI$_2$, where minima at \(\Gamma\), $M$
and off-symmetry points correspond to FM, AFM, and frustrated or spiral states, respectively. (d) Candidate materials combining incommensurate ordering vectors and weak anisotropy, indicative of spiral or canted magnetic order.}
\label{fig:simulations2}
\end{figure*}

Mapping the extracted exchange parameters onto a triangular \(J_1\)–\(J_2\)–\(J_3\) lattice model refines the microscopic picture of magnetic competition. Nearly all halides exhibit a dominant nearest-neighbor \(J_1 < 0\) (FM) interaction arising from direct $M–X–M$ overlap, opposed by an antiferromagnetic \(J_2 > 0\) term mediated by $M–X–X–M$ superexchange, while the third-neighbor coupling \(J_3\) is non-negligible and alternates in sign across compounds.
Several structures display \(|J_1/J_2| \sim 1\), signaling strong frustration and the coexistence of competing magnetic order parameters. 
This is corroborated by the DFT energetics, where FM and AFM configurations often differ by only a few tens of meV per formula unit (Figure~\ref{fig:simulations}(b)). 
These results demonstrate that the mixed region of the SOM does not arise solely from model uncertainty, but instead reflects genuine energetic competition between itinerant Stoner-like exchange and localized superexchange-driven magnetism. 
Such regime represents a physical boundary in which FM, AFM, and frustrated states coexist within a fragile balance of exchange interactions, naturally manifesting as a blurred SOM region between itinerant and localized magnetic behavior.
We next examine the coupling between spin orientation and lattice symmetry by performing SOC calculations to determine the magnetocrystalline anisotropy energy (MCAE; Figure~\ref{fig:simulations}(d), defined as the total-energy difference between magnetization along the out-of-plane (\(z\)) axis and the in-plane directions (\(x\) or \(y\)). Because long-range magnetic order is inherently unstable in strictly two-dimensional systems, SOC-induced anisotropy plays a critical role in stabilizing finite-temperature magnetism; without it, spontaneous ordering is forbidden by the Mermin–Wagner theorem. We find that FM materials generally exhibit smaller and more isotropic MCAE values, consistent with itinerant Stoner-type exchange in which delocalized electrons couple weakly to the lattice. In contrast, AFM materials display significantly larger and more anisotropic MCAE values, with pronounced differences between \(E_x\) and \(E_y\), indicative of easy-axis reorientation and strong SOC-driven orbital polarization associated with localized \(d\)-states.
Such behavior suggests that AFM systems, particularly those with broken inversion symmetry or reduced point-group symmetry, may support complex spin textures—including spin spirals or Dzyaloshinskii–Moriya (DM) canting—that extend beyond simple FM–AFM competition. These effects reflect a deeper coupling between electronic structure, lattice symmetry, and relativistic exchange interactions.
To further probe this interplay, we employ the Luttinger–Tisza (LT) method, a linearized approach that identifies the magnetic ordering wave vector \(\mathbf{q}^*\) by minimizing the Heisenberg exchange energy. By relaxing the fixed-moment constraint and diagonalizing the Fourier-transformed exchange matrix \(J(\mathbf{q})\), we determine whether FM, AFM, or incommensurate spiral order is energetically favored. 
Exchange-energy maps \(E(h,k)\) are computed as a function of the propagation vector \(\mathbf{q} = h\mathbf{b}_1 + k\mathbf{b}_2\) and are shown in Figure~\ref{fig:simulations2}(a–c) for representative materials NiCl$_2$, NiBr$_2$, and CoI$_2$.
Energetically favorable magnetic configurations are marked by minima in \(E(h,k)\): a minimum at \(\Gamma\) corresponds to FM order, at \(M\) to stripe or AFM order, and at off-symmetry points to frustrated or spiral configurations. The degree of magnetic competition is quantified using a frustration index,

\begin{equation}
f = \frac{|J_2| + |J_3|}{|J_1|},
\end{equation}

which measures the relative strength of further-neighbor interactions. Materials with \(f \ll 1\) exhibit well-defined FM or AFM order, whereas \(f \sim 1\) corresponds to broadened or displaced minima in \(E(h,k)\), indicative of frustration and a tendency toward spiral states. Combining these results with SOC calculations, we identify systems with both incommensurate LT minima (\(|\mathbf{q}^*| \neq 0\)) and weak anisotropy (\(\mathrm{MCAE} \leq 1~\mathrm{meV}\) per formula unit) as candidates. This threshold is estimated from the experimentally confirmed spin spiral in NiI$_2$ \cite{miao2025spin}, for which our calculated MCAE is 0.62 meV. We note that no universally agreed  value exists and this bound should be regarded as approximate. Only a small subset—NiCl$_2$\_2, CoBr$_2$\_3, CrO$_2$\_2, NiBr$_2$\_3, and CoCl$_2$\_1—satisfies both criteria (Figure~\ref{fig:simulations2}(d).
This trend aligns with symmetry expectations: 1T structures are centrosymmetric and therefore DM-forbidden, favoring robust FM (\(\Gamma\)) or stripe (\(M\)) order when MCAE is large. In contrast, 1H and \(P\bar{4}m2\) structures lack inversion symmetry, allowing DM interactions which, when combined with weak anisotropy, produce incommensurate LT minima consistent with canted or spiral magnetic states.
%

%
\subsection{Key findings from SEF-guided mechanisms}
Overall, the ML classification models trained on the SEF identify compounds exhibiting nontrivial magnetic behavior, rationalized in terms of symmetry, local coordination, and electronic structure. Although trained on monolayer unit cells, the SEF’s invariance to atom permutation, rotation, translation, and symmetry operations enables robust generalization to larger supercells. Candidate systems highlighted by the SOM are further validated using DFT in combination with Luttinger–Tisza analysis, confirming that the interplay between spin degrees of freedom and lattice symmetry modulates the balance between itinerant and localized exchange mechanisms. This analysis also identifies a subset of materials (e.g., NiCl$_2$, CoBr$_2$, NiBr$_2$) with potential canted or spiral magnetic states arising from competing interactions and weak anisotropy. 
%
\begin{figure}
\centering
\includegraphics[width=\columnwidth]{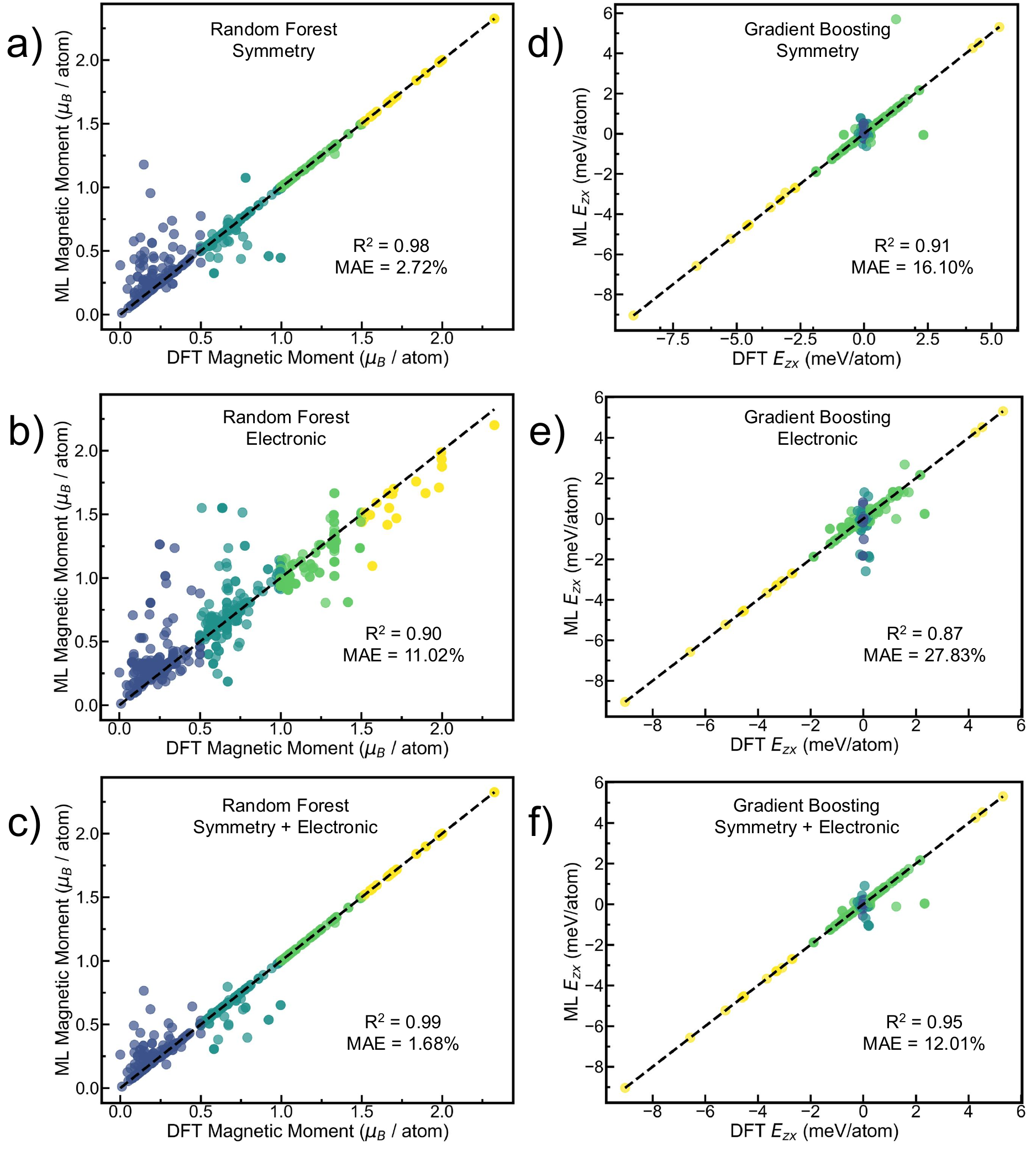}
\caption{Performance of regression models trained with (symmetry-only, electronic-only and SEF to predict magnetic moment (a-c) and magnetocrystalline anisotropy (d-f), respectively.}
\label{fig:regression}
\end{figure}

%
\begin{figure*}
\centering
\includegraphics[width=0.65\textwidth]{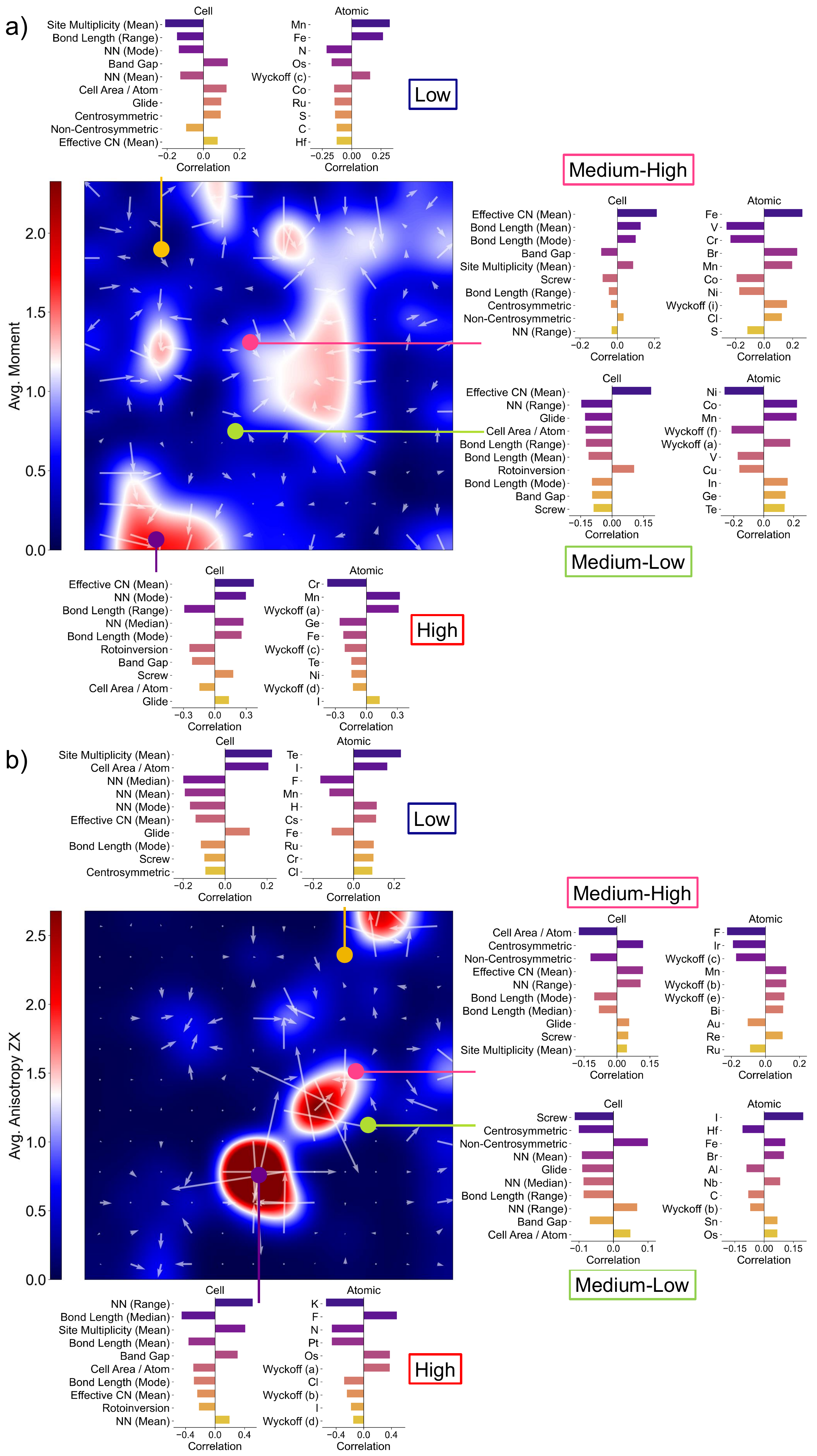}
\caption{SOM map of feature space divided into regions based on magnetic strength ($\mu_B$/atom, a) and magnetic anisotropy (meV/atom, b). To account for the presence of outliers with anomalously large MCAE values, the colormap in (b) is normalized to the 97th percentile of the distribution.}
\label{fig:SOM_reg}
\end{figure*}
%
\subsection{SEF-based regression of magnetic moments and anisotropy}
Having established the microscopic origins of the SOM-derived classifications through SEF, DFT and revealed the presence of competing magnetic states and exchange interactions, we further assess whether the learned fingerprints can capture quantitative relationships needed to distinguish itinerant, localized, and frustrated regimes. Specifically, we apply them to regression tasks aimed at predicting continuous magnetic properties – magnetic moment and MCAE (Figure~\ref{fig:regression}). Demonstrating accurate regression performance would confirm that the model encodes not only categorical distinctions between magnetic states but also the quantitative magnetic physics governing exchange strength and spin-lattice coupling.
Similar to the classification task, AFM-labeled structures are excluded due to unreliable labels, and NM systems are omitted by definition. Supervised regressors such as RFs and gradient boosting machines (GBMs) are trained to predict two continuous properties: the magnetic moment \(\mu\) (in \(\mu_B\) per atom) and the MCAE (meV per atom) along two orthogonal directions, \(\Delta E_{xz}\) and \(\Delta E_{yz}\). Models are systematically compared when trained with symmetry-based descriptors, electronic descriptors, or SEF, respectively.
To address the skewed distribution of targets, both tasks employ balanced training via binning and bootstrapping. Magnetic moments are partitioned into four regimes: low (\(0 < \mu \le 0.5\)), medium-low (\(0.5 < \mu \le 1.0\)), medium-high (\(1.0 < \mu \le 1.5\)), and high (\(\mu > 1.5\)). MCAE values, which can be positive or negative, are binned by absolute value: low (\(0 < |\Delta E_{\mathrm{MCAE}}| \le 0.025\)), medium-low (\(0.025 < |\Delta E_{\mathrm{MCAE}}| \le 0.25\)), medium-high (\(0.25 < |\Delta E_{\mathrm{MCAE}}| \le 2.5\)), and high (\(|\Delta E_{\mathrm{MCAE}}| > 2.5\)). This framework ensures uniform representation across regimes, enabling robust evaluation of model performance in capturing both subtle and large variations in ferromagnetic strength and anisotropy.
We report regression performance in terms of the percentage mean absolute error (MAE) and the coefficient of determination (\(R^2\)) in Figure~5, with random forests (RFs) applied to the magnetic moment (\(\mu\), Figure~\ref{fig:regression}(a–c) and gradient boosting machines (GBMs) applied to the MCAE along the \(xz\) direction (\(\Delta E_{xz}\), Figure~\ref{fig:regression}(d–f). Results for the \(yz\) direction (\(\Delta E_{yz}\)) are provided in {Supplementary Figure S4,S5} and exhibit nearly identical performance to the \(xz\) direction.  
Consistent with the classification task, models trained on the SEF achieve the highest accuracy. Strikingly, however, symmetry descriptors alone outperform electronic descriptors in predicting magnetic moment, indicating that while electronic fingerprints are more decisive in distinguishing FM from NM systems, symmetry fingerprints exert greater influence on the magnitude of the ordered moment once ferromagnetism is established. This is reflected in the MAE for \(\mu\), which decreases from 11.02\% (electronic only) to 2.72\% (symmetry only), and further to 1.58\% for the combined descriptor set. Correspondingly, the \(R^2\) value increases from 0.90 (electronic only) to 0.99 (combined). Comparable trends are observed for MCAE, where \(R^2\) rises from 0.87 (electronic only) to 0.95 (combined), highlighting the importance of including both descriptor types for accurate prediction of magnetic strength and anisotropy, even though symmetry descriptors dominate in influence.
\subsection{Symmetry-electronic origins of moment and anisotropy}
In Figure~\ref{fig:SOM_reg}, we extend our SOM analysis, previously applied to magnetic-state classification, to predicting magnetic strength and anisotropy, revealing a consistent physical picture. Specifically, the magnetic-moment SOM (Figure~\ref{fig:SOM_reg}(a) shows that the same structural and electronic fingerprints that promote Stoner-type itinerant ferromagnetism also delineate the high-moment basin (\(\mu \ge 1.5~\mu_B\) per atom). Materials with large, narrowly distributed bond lengths form well-defined coordination shells that enhance metal–metal orbital overlap, favoring itinerant and double-exchange mechanisms.  
Conversely, centrosymmetry, rotoinversion, screw axes, and glide planes provide symmetry-protected degeneracies which, when combined with broad bond-length distributions, suppress \(\mu\) in favor of superexchange-driven AFM or NM behavior. The medium-moment regions further illustrate this competition. The systems enriched in Fe or Mn, with broken inversion symmetry and expanded in-plane lattice parameters, systematically exhibit higher \(\mu\), whereas increasing band gap, coordination number, or number of nearest neighbors reduces it. Likewise, chalcogenides and oxides trend toward lower \(\mu\), consistent with more localized exchange pathways. Finally, the low-\(\mu\) manifold exhibits strong positive correlations with band gap, centrosymmetry, and nonsymmorphic symmetry elements (glide planes and screw axes), precisely the structural and bonding motifs expected to stabilize superexchange-dominated AFM or NM states.
The SOM analysis of the MCAE landscape (Figure~\ref{fig:SOM_reg}(b) further sharpens the mechanistic distinction between localized, SOC-driven anisotropy and itinerant Stoner magnetism. High-MCAE regions are dominated by heavy elements (e.g., Os, Bi), large band gaps, extended nearest-neighbor ranges, and centrosymmetry, with additional, though weaker, support from screw axes and increased coordination numbers. Collectively, these fingerprints indicate that strong SOC acting on relatively localized \(d\)-states governs the anisotropy in this regime.
%


%
The two intermediate basins delineate regions where structural and electronic fingerprints begin to reorganize. In the medium-high MCAE region, increased band gap, expanded nearest-neighbor ranges, and the presence of screw axes and glide planes correlate positively with MCAE, whereas large in-plane lattice area and rotoinversion suppress it. By contrast, the medium-low basin exhibits a partially reversed pattern. MCAE increases with nearest-neighbor number, cell area, and bond-length, but decreases with band gap and larger coordination numbers. Elementally, Pt and Mn contribute positively in the medium-high sector, whereas Mn, Na, and Re suppress MCAE in the medium-low sector. Finally, the low-MCAE manifold carries the fingerprints of weak SOC or itinerant magnetism: screw axes and glide planes, broad bond-length statistics, and chalcogen content (S, I) make modest positive contributions, while O, F, Ga, Ru, Rh, increased coordination numbers, and increased dispersion in the number of nearest neighbors reduce MCAE.  
The distinctions among the MCAE regions of the SOM arise from a gradual shift between two competing mechanisms: a SOC-dominated, orbital-localized regime that favors large anisotropy, and a delocalized, itinerant regime in which Stoner-type physics suppresses MCAE. The high-MCAE basin corresponds to the limit where strong SOC acting on localized \(d\)-states dominates magnetic exchange and anisotropy. Conversely, the low-anisotropy landscape reflects the itinerant, weak-SOC limit expected for Stoner-like systems.  
The intermediate regimes capture 2D materials in which competition between these mechanisms determines the MCAE. In the medium-high region, anisotropy remains largely SOC-driven, whereas the medium-low region exhibits a growing contribution from itinerancy, producing a pronounced interplay between crystal-field distortions and band dispersion. In this regime, larger bond lengths and increased variability in nearest-neighbor environments enhance local orbital asymmetry and raise MCAE, while large band gaps and elevated coordination numbers broaden the \(d\)-manifold, weaken SOC-induced level splittings, and correspondingly reduce MCAE.  
Overall, analysis of the SOM for MCAE in 2D materials demonstrates that centrosymmetric, localized, superexchange-favoring systems yield stronger magnetic anisotropy, itinerant Stoner-like physics reduces anisotropy, and intermediate regions represent a competition between the two mechanisms.
\section{Conclusions}
In this work, we introduce the symmetry-electronic fingerprint (SEF), a physics-aware materials representation that enables machine learning models to access the underlying mechanisms governing magnetism in two-dimensional materials. By explicitly encoding crystallographic symmetry, Wyckoff-site geometry, and site-resolved electronic structure directly into the representation, the SEF enables ensemble models to learn structure-property relationships grounded in the physics of exchange interactions rather than empirical correlations.
Several concrete physical insights emerge from our studies.
The Co- and Ni-based halides (e.g., NiCl$_2$, NiBr$_2$, CoI$_2$) are consistently mapped to an itinerant Stoner regime, characterized by partially delocalized $3d$ states, small band gaps, and weak magnetocrystalline anisotropy, whereas their oxide counterparts (e.g., CoO$_2$, CrO$_2$) shift toward a localized, superexchange-dominated limit with enhanced orbital localization and competing magnetic configurations. 
These regimes are not imposed a priori but emerge naturally as separable manifolds within the SEF fingerprint. Intermediate regions, occupied by materials such as NiCl$_2$ and CoI$_2$, are identified as high-uncertainty zones where Heisenberg mapping reveals $|J_1/J_2| \sim 1$, indicating strong magnetic frustration. 
The Luttinger–Tisza analysis further uncovers incommensurate or off-symmetry ordering vectors in these systems, consistent with a tendency toward noncollinear magnetism under weak anisotropy. Importantly, the SEF also exposes a systematic suppression of SOC-driven anisotropy by itinerancy, explaining why halides exhibit reduced MCAE despite comparable atomic SOC strength to their oxide counterparts. 
Crucially, the elevated predictive uncertainty localized in these mixed regimes is not incidental. Because the SEF retains direct physical meaning, uncertainty becomes a diagnostic. It acts as a compass pointing toward materials poised at the boundary between competing exchange mechanisms, where small perturbations in composition, strain, or coordination drive transitions between collinear, frustrated, and noncollinear magnetic phases. This reframes model uncertainty from a limitation into an active discovery tool, enabling targeted first-principles exploration of materials that conventional descriptors cannot distinguish from ordinary magnets.
Since the SEF is derived directly from symmetry and geometry rather than fitted to a specific chemistry, it is inherently transferable across material classes. This is the characteristic of SEF that distinguishes itself from descriptors optimized for narrow chemical spaces. More broadly, our results demonstrate that embedding physical principles directly into machine-learning representations moves beyond 
correlation and screening to enable mechanistic insight, discovery of competing phases, and hypothesis generation. Natural extensions of this representation include application to bulk magnetic systems, 
incorporation of finite-temperature spin fluctuations, and exploration of heterostructure geometries where interfacial symmetry breaking introduces additional exchange channels. In each setting, the SEF's mechanism-resolved uncertainty provides a principled starting point for targeted materials discovery.
\section {Acknowledgments}
A.G. acknowledges the hospitality of Oak Ridge National Laboratory, where this project was initialized. This work was supported by the Laboratory Directed Research and Development Program of Oak Ridge National Laboratory, managed by UT-Battelle, LLC, for the U. S. Department of Energy. This work was performed at the Center for Nanophase Materials Sciences, a U.S. Department of Energy Office of Science User Facility operated at Oak Ridge National Laboratory. ORNL is managed by UT-Battelle, LLC, for DOE under Contract No. DE-AC05-00OR22725.

\section{Author Contributions}
A.G. identified the central scientific question and developed the conceptual basis of the approach. A.F. formulated the data representation, developed the ML methods, and performed the DFT calculations. A.G. and A.F. together wrote the manuscript with helpful discussions with Z.R.F. and D.P.

\section {Conflicts of Interest}
The authors have no conflict of interest to declare.

\section {Data Availability}
The datasets utilized in this study can be found in the publicly available at ref. \cite{haastrup2018computational}.

\section {Code Availability}
All codes used in this study – such as fingerprint generation and machine learning models – will be made available upon acceptance.

\bibliography{references}

\end{document}